\documentclass[aps,prd,showpacs,onecolumn,preprintnumbers,notitlepage,
nofootinbib,amsfont,amssymb,10pt,singlespacing]{revtex4-1}
\usepackage{amsmath}
\usepackage{bm}
\usepackage{times}
\usepackage{braket}
\usepackage{color}
\usepackage{epsfig}
\usepackage{slashed}
\usepackage{hyperref}

\newcommand{\beq}{\begin{eqnarray}}
\newcommand{\eeq}{\end{eqnarray}}

\newcommand{\acp}{ A_{\rm CP}}

\def\lsim{ {\ \lower-1.2pt\vbox{\hbox{\rlap{$<$}\lower6pt\vbox{\hbox{$\sim$}
}}}\ } }
\def\gsim{ {\ \lower-1.2pt\vbox{\hbox{\rlap{$>$}\lower6pt\vbox{\hbox{$\sim$}
}}}\ } }
\def \epjc{ Eur. Phys. J. C }

\def \jpg{  J. Phys. G }

\def \plb{  Phys. Lett. B }

\def \prd{  Phys. Rev. D }

\def \zpc{  Z. Phys. C  }
\def \jhep{ J. High Energy Phys.  }

\definecolor{Red}{rgb}{1.,0.,0.}

\definecolor{Blue}{rgb}{0.,0.,1.}

\definecolor{nicered}{rgb}{0.7,0.1,0.1}
\definecolor{nicegreen}{rgb}{0.1,0.5,0.1}

\hypersetup{colorlinks,citecolor=nicegreen,linkcolor=nicered}

\begin{document}

\title{Resolving the $B \to K \pi$ puzzle by Glauber-gluon effects}

\author{Xin~Liu}
\email[Electronic address: ]{liuxin@jsnu.edu.cn}
\affiliation{School of Physics and Electronic Engineering,
Jiangsu Normal University, Xuzhou, Jiangsu 221116, People's Republic of China}

\author{Hsiang-nan~Li}
\email[Electronic address: ]{hnli@phys.sinica.edu.tw}
\affiliation{ Institute of Physics, Academia Sinica, Taipei, Taiwan 115, Republic of China}

\author{Zhen-Jun~Xiao}
\email[Electronic address: ]{xiaozhenjun@njnu.edu.cn}
\affiliation{Department of Physics and Institute of Theoretical
Physics, Nanjing Normal University, Nanjing, Jiangsu 210023,
People's Republic of China}

\date{\today}

\begin{abstract}

We extend the perturbative QCD formalism including
the Glauber gluons, which has been
shown to accommodate the measured $B \to \pi \pi$ and
$B^0\to\rho^0 \rho^0$ branching ratios simultaneously, to the analysis
of the $B \to K \pi$ and $K \bar K$ decays. It is observed that the
convolution of the universal Glauber phase factors with the
transverse-momentum-dependent kaon wave function reveals weaker (stronger)
Glauber effects than in the pion ($\rho$ meson) case as expected.
Our predictions for the branching ratios and the direct CP asymmetries
of the $B \to K \pi$ and $K \bar K$ modes at next-to-leading-order accuracy
agree well with data. In particular, the
predicted difference of the $B^\pm \to K^\pm \pi^0$ and $B^0 \to K^\pm \pi^\mp$
direct CP asymmetries, $\Delta A_{K\pi}\equiv \acp^{\rm dir}(K^\pm \pi^0)[0.021 \pm 0.016]
- \acp^{\rm dir}(K^\pm \pi^\mp)[-0.081 \pm 0.017] = 0.102 \pm 0.023$,
is consistent with the measured $\Delta  A_{K\pi} = 0.119 \pm 0.022$ within uncertainties,
and the known $B \to K\pi$ puzzle is resolved. The above
$B \to \pi \pi$, $K\pi$ and $K \bar K$ studies confirm
that the Glauber gluons associated with pseudo-Nambu-Goldstone bosons enhance
the color-suppressed tree amplitude significantly, but have a small
impact on other topological amplitudes.

\end{abstract}

\pacs{13.25.Hw, 12.38.Bx, 11.10.Hi}

\maketitle

Several long-standing puzzles in two-body charmless hadronic $B$-meson decays,
which might reveal new physics signals, have motivated thorough investigations
of heavy-quark decay dynamics. Among them, we have carefully examined the
$B\to\pi\pi$ puzzle~\cite{Li:2009wba,Li:2014haa,Liu:2015sra} 
that originates from the contradiction between the
theoretically small and experimentally large $B^0 \to \pi^0 \pi^0$ branching
ratios. It is generally believed that the enhancement of the color-suppressed
tree-amplitude $C$ resolves the $B\to\pi\pi$
puzzle~\cite{Charng:2003iy,Charng:2004ed,Buras:2003dj,CGRS}.
We then observed in~\cite{Li:2009wba,Li:2014haa} that the Glauber gluons,
resulting in a nonperturbative strong phase in the $k_T$ factorization
theorem, could provide such an enhancement mechanism: the additional Glauber phase
turns the destructive interference between the spectator diagrams in the
$B\to\pi\pi$ decays into a constructive one, and further increases the amplitude
$C$ on top of the contribution from next-to-leading-order (NLO) vertex
corrections~\cite{Li:2005kt}. The remaining challenge resides in understanding
why the Glauber-gluon effect is significant in the $B\to\pi\pi$, but not
$B\to\rho\rho$, decays, for which theoretical predictions agree well with the
data~\cite{LM06}. We have speculated~\cite{Li:2009wba,Li:2014haa} that the dual
role of a pion as a pseudo-Nambu-Goldstone (NG) boson and as a $q\bar q$ bound
state~\cite{NS08} may account for its difference from a $\rho$ meson.

The Glauber phases associated with a pion were treated as free parameters,
and those associated with a $\rho$ meson were assumed to vanish in the earlier
studies of two-body charmless hadronic $B$-meson decays~\cite{Li:2009wba,Li:2014haa}.
Recently, we derived a perturbative QCD (PQCD) factorization formalism, which
contains convolution of transverse-momentum-dependent (TMD) hadron wave functions
with universal Glauber phase factors~\cite{Liu:2015sra}. The fitting to the relevant
$B$-meson transition form factors indicated that the TMD pion ($\rho$ meson) wave
function exhibits a weak (strong) falloff in parton transverse momentum $k_T$, a
feature in accordance with the special role of a pion, which requires a tighter spatial
distribution of its leading Fock state relative to higher Fock states~\cite{NS08}
in the conjugate $b$ space. Simply parametrizing the Glauber phase as an sinusoidal
function in the $b$ space, we showed quantitatively that the leading Fock state of
a pion may be tight enough to explore the Glauber effect from the oscillatory
phase factor on the enhancement of the amplitude $C$, while the leading Fock state
of a $\rho$ meson is not. Our detailed analysis in~\cite{Liu:2015sra} confirmed
that the convolution of the universal Glauber phase factors with the TMD pion ($\rho$ meson)
wave function leads to large (moderate) modification of the $B^0\to\pi^0\pi^0$
($B^0\to\rho^0\rho^0$) branching ratio. Namely, the $B\to\pi\pi$ puzzle could be resolved
under the stringent constraint from the $B^0 \to \rho^0 \rho^0$ data.

Encouraged by the success of the PQCD approach with the Glauber effects
on accommodating the measured $B \to \pi \pi$ and $B^0\to\rho^0 \rho^0$ branching
ratios simultaneously, we extend it to the investigation of the known $B \to K \pi$
puzzle~\cite{Kpi_puzzle}: the direct CP asymmetries of the $B^0 \to K^\pm
\pi^\mp$ and $B^\pm \to K^\pm \pi^0$ modes are expected to be
roughly equal, but the data~\footnote{The most
precise measurement of $\acp^{\rm dir}(B^0 \to K^\pm \pi^\mp)$ to date was performed by
the Large Hadron Collider-beauty (LHCb) Collaboration, giving $-0.080
\pm 0.007({\rm stat}) \pm 0.003({\rm syst})$~\cite{Aaij:2013iua}.}\cite{Agashe:2014kda,Amhis:2014hma},
\beq
     \acp^{\rm dir}(B^0 \to K^\pm \pi^\mp) &=& -0.082 \pm 0.006\;, \\
     \acp^{\rm dir}(B^\pm \to K^\pm \pi^0) &=& +0.037 \pm 0.021\;,
\eeq
imply the deviation of their difference
$\Delta A_{K\pi}\equiv\acp^{\rm dir}(K^\pm \pi^0)-\acp^{\rm dir}(K^\pm \pi^\mp)
= 0.119 \pm 0.022$ from zero at $5\sigma$ level. The $B \to K\pi$ decays
have been discussed in the NLO PQCD approach in
Refs.~\cite{Li:2005kt,Wei:2014ioa,Li:2014haa}, whose
predictions for the branching ratios match the
data within theoretical errors in general, but those for $\Delta A_{K\pi}$
still do not. It is then worthwhile to examine whether the $B \to K \pi$
puzzle can be resolved in the same formalism with the universal Glauber phase
factors.

It will be demonstrated that the TMD kaon wave function, extracted from the
$B\to K$ transition form factor, exhibits a stronger (weaker) falloff in the
parton transverse momentum $k_T$ than the TMD pion ($\rho$ meson) wave function
does. That is, the leading $q\bar q$ Fock state of a kaon, which is also a
pseudo-NG boson, is not as tight as of a pion in the spatial distribution.
This difference can be understood as a consequence of SU(3) symmetry breaking.
The convolution of the universal Glauber phase factors with the TMD kaon wave
function then reveals weaker (stronger) Glauber effects than in the pion
($\rho$ meson) case. For a complete analysis of the Glauber effects
associated with the kaon, we also consider the $B \to K \bar K$
modes, for which NLO PQCD predictions have not yet been available.
It will be seen at NLO accuracy that our predictions for the branching
ratios and the direct CP asymmetries of the $B \to K \pi$
and $K \bar K$ decays all agree well with data. In particular, the
predicted difference $\Delta A_{K\pi} \equiv
\acp^{\rm dir}(K^\pm \pi^0)[0.021 \pm 0.016]
- \acp^{\rm dir}(K^\pm \pi^\mp)[-0.081 \pm 0.017] = 0.102 \pm 0.023$
is consistent with the measured $\Delta  A_{K\pi} = 0.119 \pm 0.022$ within uncertainties,
and the $B \to K\pi$ puzzle is resolved. Overall, the impact
of the Glauber effects on the $B \to K \pi$ modes is more significant than on
the $B \to K \bar K$ ones, as expected, since the latter do not involve the
amplitude $C$.

Following Refs.~\cite{Huang80,BHP80,YXM07}, we write the intrinsic $k_T$-dependent
kaon wave function as
\begin{eqnarray}
\phi_K(x,{\bf k}_T)=\frac{\pi}{2\beta_K^2}
\exp\left(-\frac{{\cal M}^2}{8\beta_K^2}\right)\frac{\phi_K(x)}{x(1-x)},
\label{mod3}
\end{eqnarray}
where $x$ is the parton momentum fraction of the light quark, $\beta_K$ is a shape parameter,
and $\phi_K(x)$ represents the standard twist-2 and twist-3 light-cone distribution
amplitudes. The first (second) argument $k_T$ in the factor
\begin{eqnarray}
{\cal M}^2=\frac{k_T^2+m_q^2}{x}+\frac{k_T^2+m_s^2}{1-x},\label{mm}\;,
\end{eqnarray}
denotes the transverse momentum carried by the light (strange) quark, with $m_{q(s)}$
standing for the light (strange) quark mass.
The two-argument kaon wave function~\cite{CL09} to be convoluted with the
Glauber phase factors in the PQCD framework~\cite{Liu:2015sra} is then given by
\begin{eqnarray}
\bar\phi_K(x,{\bf b}',{\bf b})&\equiv&
\int \frac{d^2 {\bf k}'_T}{(2\pi)^2}\frac{d^2 {\bf k}_T}{(2\pi)^2}\exp(-i{\bf k}'_{T}
\cdot {\bf b}')\exp(-i{\bf k}_{T}\cdot {\bf b})\phi_K(x,{\bf k}'_{T},{\bf k}_{T}),\nonumber\\
&=&\frac{2\beta_K^2}{\pi}\exp{\left[-\frac{1}{8\beta_K^2}(\frac{m_q^2}{x}+\frac{m_s^2}{1-x})\right]} \exp\left[-2\beta_K^2xb^{\prime
2}-2\beta_K^2(1-x)b^{2}\right]\phi_K(x) .\label{6}
\end{eqnarray}
In the above expression $b^{\prime}$ ($b$) labels the transverse coordinate of the light (strange)
quark of a kaon.

The shape parameter $\beta_K$ is constrained neither theoretically nor
experimentally. Therefore, we determine it in the way the same
as determining the $\rho$-meson shape parameter~\cite{Liu:2015sra}, namely,
by fitting to the $B \to K$ transition form factor $F_0^{B\to K}(0)\sim 0.34$,
which is better known in the literature. The value
$\beta_K \sim 0.25$ GeV is obtained, implying a stronger (weaker) falloff in
$k_T$ compared to the pion ($\rho$-meson) wave function with $\beta_\pi \sim 0.40$
($\beta_\rho \sim 0.13$) GeV~\cite{Liu:2015sra}.
That is, a kaon demands a broader (tighter) spatial distribution in the $b$ space
for the leading $q\bar q$ Fock state than the pion ($\rho$ meson) does. It is then
expected that the TMD kaon wave function can partially probe the oscillatory phase of
the universal Glauber factor and reveal the Glauber effect in between the pion and
$\rho$ meson cases. The underlying reason is that a kaon, despite 
also being a NG boson, has less tension to its 
other role as a $q\bar q$
bound state due to the larger mass, i.e., due to SU(3) symmetry breaking. We will verify
this observation by explicitly evaluating the Glauber effects on the
color-suppressed tree amplitude later.

We take the distribution amplitudes
\begin{eqnarray}
\phi_{K}^A(x)&=&\frac{6f_K}{2\sqrt{2N_c}}x(1-x)\bigg[1+3 a_1^K (2x-1)+
\frac{3}{2} a_2^{K} \bigg( 5(2x-1)^2  - 1 \bigg)+ \frac{15}{8} a_4^{K}
\bigg(1- 14(2x-1)^2 +4 (2x -1)^4\bigg)\bigg],\nonumber\\
\phi_{K}^P(x)&=&\frac{f_{K}}{2\sqrt{2N_c}}\, \bigg[ 1
+\frac{1}{2} \left(30\eta_3 -\frac{5}{2}\rho_{K}^2\right) \bigg(3 (2x-1)^2 -1\bigg) \nonumber\\ &
& \hspace{35mm} -\, \frac{3}{8}\left\{ \eta_3\omega_3 +
\frac{9}{20}\rho_{K}^2(1+6a_2^{K}) \right\} \bigg(3- 30 (2x-1)^2 +35 (2x-1)^4\bigg)
\bigg]\;,\\
\phi_{K}^T(x)&=&\frac{f_{K}}{2\sqrt{2N_c}}\,
(1-2x)\bigg[ 1 + 6\left(5\eta_3 -\frac{1}{2}\eta_3\omega_3 -
\frac{7}{20}
      \rho_{K}^2 - \frac{3}{5}\rho_{K}^2 a_2^{K} \right)
(1-10x+10x^2) \bigg],
\end{eqnarray}
for a kaon~\cite{Ball99:Pseudoscalar}, whose parametrization is the same as
for a pion in Ref.~\cite{Liu:2015sra} but with different hadronic parameters.
Here we adopt the light $u,d$ quark mass $m_q= 7.5$ MeV,
the strange quark mass $m_s =130$ MeV~\cite{Agashe:2014kda}, the kaon decay constant
$f_K = 0.16$ GeV, the Gegenbauer moments $a_1^K = 0.17$ and $a_2^K = 0.115$~\cite{Li:2005kt},
the coefficients $a_4^K=-0.015$, $\eta_3 =0.015$, $\omega_3 =-3$, and the factor
$\rho_K = m_K/m_0^K$ with the kaon mass $m_K=0.496$ GeV and the chiral mass
$m_0^K = 1.7$ GeV~\cite{Li:2005kt}. The choice of the quark masses is not crucial, and their variation can be compensated by that of the Gegenbauer moments. The hadronic parameters
associated with a pion are the same as in Ref.~\cite{Liu:2015sra}.
We have checked that the NLO results for the $B \to K \pi$ decays~\cite{Li:2005kt} in
the PQCD approach without the Glauber effects are reproduced with the above
TMD wave functions and parameters. It is noticed that the
$B \to \pi$ and $B \to K$ transition form factors
$F_0^{B \to \pi}(0)$ and $F_0^{B\to K}(0)$ in Ref.~\cite{Wei:2014ioa},
which include the NLO contributions, are
smaller than those in Refs.~\cite{Li:2014haa,Li:2005kt}.

The parametrized Glauber phase factor $S({\bf b})=r\pi\sin(p b)$, in which
the parameters $r=0.60$ and $p=0.544$ GeV govern the magnitude and the
frequency of the oscillation, also remain
the same as in the analysis of the $B\to \pi \pi$ and $\rho \rho$
decays~\cite{Liu:2015sra} because of their universality for two-body charmless
hadronic $B$-meson decays. The oscillatory behavior can be understood via
the Fourier transformation of the nonperturbative Glauber gluon propagator,
$\int^\Lambda_0 d^2l_T\exp(-i{\bf l}_{T}\cdot {\bf b})/(l_T^2+m_g^2)$, with $m_g$
being a gluon mass. Choosing the cutoff for the loop momentum,
$\Lambda\sim 0.5$ GeV, which is reasonable for collecting soft contribution,
roughly yields the period $p$ obtained in~\cite{Liu:2015sra}. The gluon mass
$m_g$, together with the coefficient in the associated loop correction,
such as the strong coupling in the nonperturbative region,
control the magnitude of the oscillation.
The $b\to 0$ limit corresponds to the integration over the transverse momentum,
namely, to the collinear factorization theorem. It has been known
that the color-suppressed tree amplitude is dominated by the
contribution from an energetic spectator quark in the collinear factorization
for two-body charmless hadronic $B$-meson decays~\cite{Beneke:2000ry}.
According to the discussion in~\cite{Li:2009wba,Li:2014haa}, the Glauber
region is not pinched as a spectator quark becomes energetic, implying
the vanishing of the Glauber phase in the $b\to 0$ limit. The above reasoning
explains the parametrization $S({\bf b})=r\pi\sin(p b)$ in Ref.~\cite{Liu:2015sra}.

\begin{table}[t]
\caption{Branching ratios of the $B \to K \pi$ decays from the NLO PQCD formalism in units of
$10^{-5}$, in which NLO (NLOG) denotes the results without (with) the Glauber
effects.}\label{br1}
\begin{center}
\begin{tabular}{cccc}
\hline\hline Modes & Data \cite{Agashe:2014kda,Amhis:2014hma} 
& NLO & NLOG
\\
\hline $B^0 \to K^\pm \pi^\mp$ & $ \phantom{0} 1.96 \pm 0.05 $ &
 $\phantom{0}2.33^{+0.74}_{-0.52}(\omega_B)^{+0.12}_{-0.11}(a^K)^{+0.21}_{-0.20}(a^\pi)$&
 $\phantom{0}2.17^{+0.71}_{-0.49}(\omega_B)^{+0.11}_{-0.10}(a^K)^{+0.17}_{-0.16}(a^\pi)$
\\
$B^\pm \to K^\pm \pi^0$ & $ \phantom{0} 1.29 \pm 0.05 $ &
 $\phantom{0}1.53^{+0.50}_{-0.35}(\omega_B)^{+0.08}_{-0.07}(a^K)^{+0.12}_{-0.12}(a^\pi)$&
 $\phantom{0}1.40^{+0.46}_{-0.33}(\omega_B)^{+0.06}_{-0.06}(a^K)^{+0.10}_{-0.10}(a^\pi)$
\\
$B^\pm \to \pi^\pm K^0$ & $ \phantom{0} 2.37 \pm 0.08 $ &
 $\phantom{0}2.72^{+0.88}_{-0.61}(\omega_B)^{+0.15}_{-0.13}(a^K)^{+0.25}_{-0.24}(a^\pi)$&
 $\phantom{0}2.41^{+0.80}_{-0.56}(\omega_B)^{+0.11}_{-0.11}(a^K)^{+0.17}_{-0.17}(a^\pi)$
\\
$B^0 \to K^0 \pi^0$ & $ \phantom{0} 0.99 \pm 0.05 $ &
 $\phantom{0}1.02^{+0.32}_{-0.22}(\omega_B)^{+0.05}_{-0.05}(a^K)^{+0.11}_{-0.10}(a^\pi)$&
 $\phantom{0}0.93^{+0.30}_{-0.21}(\omega_B)^{+0.06}_{-0.05}(a^K)^{+0.08}_{-0.07}(a^\pi)$
\\
\hline\hline
\end{tabular}
\end{center}
\end{table}

\begin{table}[t]
\caption{Direct CP asymmetries of the $B \to K \pi$ decays from the NLO PQCD formalism, in which NLO (NLOG)
denotes the results without (with) the Glauber
effects.}\label{dcp1}
\begin{center}
\begin{tabular}{cccc}
\hline\hline Modes & Data \cite{Agashe:2014kda,Amhis:2014hma} 
& NLO & NLOG
\\
\hline $B^0 \to K^\pm \pi^\mp$ & $ \phantom{0} -0.082 \pm 0.006 $ &
 $\phantom{0}-0.076^{+0.008}_{-0.009}(\omega_B)^{+0.013}_{-0.013}(a^K)^{+0.007}_{-0.007}(a^\pi)$&
 $\phantom{0}-0.081^{+0.009}_{-0.009}(\omega_B)^{+0.011}_{-0.011}(a^K)^{+0.010}_{-0.009}(a^\pi)$
\\
$B^\pm \to K^\pm \pi^0$ & $ \phantom{0} +0.037 \pm 0.021 $ &
 $\phantom{0}-0.008^{+0.008}_{-0.009}(\omega_B)^{+0.009}_{-0.009}(a^K)^{+0.006}_{-0.006}(a^\pi)$&
 $\phantom{0}+0.021^{+0.008}_{-0.008}(\omega_B)^{+0.003}_{-0.004}(a^K)^{+0.014}_{-0.013}(a^\pi)$
\\
$B^\pm \to \pi^\pm K_S^0$ & $ \phantom{0} -0.017 \pm 0.016 $ &
 $\phantom{0}+0.003^{+0.001}_{-0.000}(\omega_B)^{+0.001}_{-0.001}(a^K)^{+0.000}_{-0.000}(a^\pi)$&
 $\phantom{0}+0.004^{+0.000}_{-0.001}(\omega_B)^{+0.001}_{-0.002}(a^K)^{+0.000}_{-0.000}(a^\pi)$
\\
$B^0 \to K_S^0 \pi^0$ & $ \phantom{0} 0.00 \pm 0.13 $ &
 $\phantom{0}-0.056^{+0.001}_{-0.001}(\omega_B)^{+0.004}_{-0.004}(a^K)^{+0.000}_{-0.001}(a^\pi)$&
 $\phantom{0}-0.089^{+0.001}_{-0.000}(\omega_B)^{+0.013}_{-0.009}(a^K)^{+0.005}_{-0.005}(a^\pi)$
\\
\hline\hline
\end{tabular}
\end{center}
\end{table}

The explicit factorization formulas for a general $B\to M_1M_2$ decay
containing the convolution with the universal Glauber phase factors
can be found in Ref.~\cite{Liu:2015sra}.
The results for all the considered quantities
are listed in Tables~\ref{br1} and~\ref{dcp1}.
Generally speaking, the NLO PQCD predictions for the $B \to K\pi$ branching ratios
without the Glauber effects
match the data within theoretical errors. Taking into account
the Glauber effects, one sees that the predicted branching
ratios, labeled by NLOG, are reduced by around 10\% as indicated in Table~\ref{br1}.
The moderate dependence of these branching ratios on the Glauber effects
is attributed to the fact that they are dominated by the penguin contributions,
and not sensitive to the color-suppressed tree amplitude $C$. However, the direct
CP asymmetry of the $B^\pm \to K^\pm \pi^0$ mode is sensitive to $C$, which may
be modified significantly by the Glauber effects. Table~\ref{dcp1} shows that
$\acp^{\rm dir}(K^\pm \pi^0)$ flips sign, changing from the central value
$-0.008$ to $+0.021$. The NLOG results in Tables~\ref{br1} and~\ref{dcp1},
agreeing well with the data within errors, represent the NLO PQCD predictions,
which best accommodate the data to date.

\begin{figure}[htb]
\begin{center}
\begin{tabular}{c}
\includegraphics[height=3.3cm]{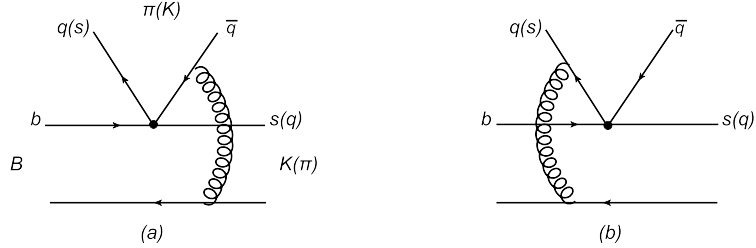}
\end{tabular}
\caption{LO spectator diagrams for the $B \to K\pi$ decays corresponding to the pion
or kaon emission from the weak vertex.
}\label{fig1}
\end{center}
\end{figure}

To have a clear idea of the Glauber effects on the $B \to K \pi$ decays, we
present the amplitudes (in units of $10^{-2}$
${\rm GeV}^{3}$) from the two leading-order (LO) spectator diagrams in Figs.~\ref{fig1}(a)
and~\ref{fig1}(b) associated with the four-fermion operator $O_2$,
\begin{eqnarray}
{\cal A}_{a,b}(B^\pm\to \pi^0 K^\pm)&=& \left\{ \begin{array}{lll}
-16.71 +i 13.71, & \ \  10.85 - i 9.96,\ \ & {\rm (NLO)}, \\
-12.57 + i 10.80,& \ \   -9.96 + i 5.85,\ \ &  {\rm (NLOG)},\\
\end{array} \right.\label{Kpi-P}
\end{eqnarray}
for the pion emission from the weak vertex, and the amplitudes
associated with the four-fermion operator $O_1$,
\begin{eqnarray}
{\cal A}_{a,b}(B^\pm\to K^\pm\pi^0)&=& \left\{ \begin{array}{lll}
4.55 -i 3.98,& \ \  -3.37+i 2.25,\ \   & {\rm (NLO)}, \\
3.59-i 3.23,& \ \ \ \ \ 3.14-i 0.90 ,\ \ &  {\rm (NLOG)},\\
\end{array} \right.\label{Kpi-K}
\end{eqnarray}
for the kaon emission. Here ${\cal A}_{a}$ and ${\cal A}_{b}$ denote the
amplitudes corresponding to the spectator diagrams with the hard gluons
attaching to the valence antiquark and the valence quark, respectively.
As exhibited in Eqs.~(\ref{Kpi-P}) and~(\ref{Kpi-K}), the destructive
interferences between the two spectator diagrams have become constructive ones
under the Glauber effects, due to the significant modification
of the amplitudes ${\cal A}_{b}$ with a sign flip. The summation of the amplitudes
changes from $6.96e^{i2.57}$ $(2.09e^{-i0.97})\times 10^{-2}$ GeV$^3$
to $28.01e^{i2.51}$ $(7.90e^{-i0.55})\times 10^{-2}$ GeV$^3$ after including
the Glauber effects associated with the TMD pion (kaon) wave functions
in the NLO PQCD approach. In particular, the modification in Eq.~(\ref{Kpi-P}),
which contributes to the color-suppressed tree amplitude,
results in the rotation of the total tree amplitude by a strong phase, and
turns the previous negative direct CP asymmetry of the $B^\pm \to K^\pm \pi^0$
modes into a positive value.

To examine the similarity between the kaon and the
pion from the viewpoint of Glauber gluons, we calculate the
spectator amplitudes, which contain only the Glauber phase factor $S_{e2}$ associated with
the emitted meson $M_2$ in the $B \to M_1 M_2$ decays.
The ratios of the ${\rm NLOG-S_{e2}}$ amplitudes over the NLO amplitudes in magnitude,
\beq
 R_\pi &\equiv& \frac{|{\cal A}_a(\pi^0 K^\pm)+{\cal A}_b(\pi^0 K^\pm)|_{\rm NLOG-S_{e2}}}{|{\cal A}_a(\pi^0 K^\pm)+{\cal A}_b(\pi^0 K^\pm)|_{\rm NLO}} \approx 4.69\;,\\
 R_K &\equiv& \frac{|{\cal A}_{a}(K^\pm \pi^0)+{\cal A}_{b}(K^\pm \pi^0)|_{\rm NLOG-S_{e2}}}{|{\cal A}_{a}(K^\pm \pi^0)+{\cal A}_{b}(K^\pm \pi^0)|_{\rm NLO}} \approx 3.90\;,
\eeq
indicate that a kaon reveals weaker Glauber enhancement than a pion does.
Besides, both the pion and the kaon reveal the Glauber effects more dramatically
than that given in Eq.~(35) of Ref.~\cite{Liu:2015sra} for a $\rho$ meson.
As stated before, a kaon is also a pseudo-NG boson like a pion,
but with non-negligible SU(3) symmetry breaking. It should be stressed that the
direct CP asymmetry of the $B^\pm\to K^\pm \pi^0$ modes predicted by the NLO PQCD
formalism will not flip sign if only the Glauber phase factor $S_{e2}$ is considered.
This fact confirms the effect of the Glauber phase
factor $S_{e1}$ associated with the meson $M_1$, which gives an additional strong phase
to the color-suppressed tree amplitude having been enhanced by $S_{e2}$~\cite{Li:2014haa}.
That is, both Glauber phase factors are crucial for resolving the $B\to K\pi$ puzzle.

%

\begin{table}[t]
\caption{Same as Table~\ref{br1} but for the $B \to K \bar K$ decays (in units of
$10^{-6}$).}\label{br2}
\begin{center}
\begin{tabular}{cccc}
\hline\hline Modes & Data \cite{Aaij:2013fja,Agashe:2014kda,Amhis:2014hma} 
& NLO & NLOG
\\
\hline $B^\pm \to K^\pm \bar K^0$ & $ \phantom{0} 1.52 \pm 0.22$~\footnote{This
is the very recent measurement reported by the LHCb Collaboration~\cite{Aaij:2013fja}, which is
comparable with $1.64 \pm 0.45$ by the BABAR Collaboration~\cite{Aubert:2006gm}
and a bit larger than $1.11 \pm 0.20$ by the Belle Collaboration~\cite{Duh:2012ie}.}  &
 $\phantom{0}2.45^{+0.83}_{-0.58}(\omega_B)^{+0.17}_{-0.17}(a^K)$&
 $\phantom{0}2.27^{+0.79}_{-0.54}(\omega_B)^{+0.17}_{-0.14}(a^K)$
\\
$B^0 \to K^0 \bar K^0$ & $ \phantom{0} 1.21 \pm 0.16 $ &
 $\phantom{0}2.19^{+0.77}_{-0.54}(\omega_B)^{+0.09}_{-0.09}(a^K)$&
 $\phantom{0}2.02^{+0.72}_{-0.50}(\omega_B)^{+0.08}_{-0.08}(a^K)$
\\
\hline\hline
\end{tabular}
\end{center}
\end{table}

\begin{table}[t]
\caption{Same as Table~\ref{dcp1} but for the $B \to K \bar K$ decays.}\label{dcp2}
\begin{center}
\begin{tabular}{cccc}
\hline\hline Modes & Data \cite{Aaij:2013fja,Agashe:2014kda,Amhis:2014hma} 
& NLO & NLOG
\\
\hline $B^\pm \to K^\pm \bar K_S^0$ & $ \phantom{0} -0.21 \pm 0.14 $ &
 $\phantom{0}-0.03^{+0.01}_{-0.01}(\omega_B)^{+0.02}_{-0.02}(a^K)$&
 $\phantom{0}-0.03^{+0.01}_{-0.01}(\omega_B)^{+0.02}_{-0.02}(a^K)$
\\
$B^0 \to K_S^0 \bar K_S^0$ & $ \phantom{0} 0.0 \pm 0.4 $ &
 $\phantom{0}-0.09^{+0.00}_{-0.00}(\omega_B)^{+0.01}_{-0.01}(a^K)$&
 $\phantom{0}-0.09^{+0.00}_{-0.00}(\omega_B)^{+0.00}_{-0.00}(a^K)$
\\
\hline\hline
\end{tabular}
\end{center}
\end{table}

The $B \to K \bar K$ decays were investigated
in the LO PQCD approach~\cite{Chen:2000ih} more than a decade ago.
The NLO results for these modes in Tables~\ref{br2}
and~\ref{dcp2} with and without the Glauber effects are derived
for the first time. It has been known that there is no amplitude
$C$ in the $B \to K \bar K$ decays, only the spectator amplitudes induced by the
penguin operators, which do not exhibit strong cancellation at LO.
Taking the $B^+ \to K^+ \bar K^0$ mode as an example, one observes that
the predicted branching ratio decreases by a few percent,
from $2.45^{+0.85}_{-0.60} \times 10^{-6}$ in the NLO PQCD formalism to
the NLOG one $2.27^{+0.81}_{-0.56} \times 10^{-6}$, while the
direct CP asymmetry remains unchanged, namely, $-0.03 \pm 0.02$. The small
reduction of the decay rate and the invariance of the direct CP asymmetry
under the Glauber effects are expected due to the absence of the amplitude $C$ here.
Within large theoretical errors, the branching ratios and the direct CP
asymmetries of the $B \to K \bar K$ decays predicted in the NLOG PQCD framework
match the existing measurements generally.

In summary, we have estimated the Glauber-gluon effects in the $B\to K\pi$
and $K \bar K$ decays in the PQCD approach at NLO level by convoluting the universal
Glauber phase factors with the TMD meson wave functions. It has been pointed
out that the kaon behaves more like the pion but with a broader spatial distribution
of the leading Fock state in the $b$ space conjugate to the parton transverse momentum
$k_T$, which causes smaller enhancement
of the color-suppressed tree amplitude $C$ with the kaon emission, relative to
the pion emission. The $B \to K \pi$ branching ratios, being
insensitive to $C$, are reduced by only around 10\%. However,
the Glauber phase factors lead to the enhancement and the rotation of the tree amplitude
by a strong phase in the $B^\pm \to K^\pm \pi^0$ modes, rendering the predicted direct
CP asymmetry consistent with the data. All the branching
ratios and the direct CP asymmetries in the $B \to K \pi$ decays then agree well with the
measurements within errors, and the $B \to K\pi$ puzzle is
resolved in the NLO PQCD formalism with the Glauber effects.
The $B \to K \bar K$ decays, which do not involve the
amplitude $C$, were also investigated in the same framework. As expected,
the Glauber gluons associated with the TMD kaon wave function do not make
a sizable impact on these modes, and the NLO PQCD predictions have matched the data
of the branching ratios and direct CP asymmetries generally.
Our recent analysis of the $B \to \pi \pi$, $\rho^0 \rho^0$, $K \pi$, and $K \bar K$
decays seems to indicate that
the convolution of the universal Glauber phase factors with different TMD meson
wave functions can generate appropriate enhancements and rotations of the
amplitude $C$ for resolving the $B\to\pi\pi$ and $K\pi$ puzzles simultaneously. We have
observed a significant impact on the amplitude $C$ from the Glauber gluons
associated with pseudo-NG bosons, and might have found a plausible dynamical origin of the
additional strong phases required by the data of the above two-body charmless
hadronic $B$-meson decays.

\begin{acknowledgments}

We thank U. Nierste and Y.M. Wang for useful discussions. This work was supported
in part by the Ministry of Science and Technology of the Republic of China under Grant No.
NSC-104-2112-M-001-037-MY3, by the National Science Foundation
of China under Grants No.~11205072 and No.~11235005, and by the Priority Academic
Program Development of Jiangsu Higher Education Institutions (PAPD).

\end{acknowledgments}

\end{document}